\documentclass[12pt,prd,aps,amssymb,amsmath,tightenlines,showpacs]{article}
\usepackage[utf8]{inputenc}
\usepackage[a4paper,top=3cm,bottom=3cm,left=1.5cm,right=1.5cm]{geometry}
\usepackage{amssymb}
\usepackage{amsmath}
\usepackage{amsthm}
\usepackage{physics}
\usepackage{mathtools}
\usepackage{tipa}
\usepackage{latexsym} 
\usepackage{graphicx}
\usepackage{slashed}
\usepackage{bbold}
\usepackage{cancel}
\usepackage{comment}
\usepackage{color}
\usepackage{graphicx,epsfig,color}
\usepackage[dvipsnames]{xcolor}
\usepackage{comment}
\usepackage{cite}
\usepackage{tikz}
\usepackage[compat=1.1.0]{tikz-feynman}
\usepackage{caption}
\usepackage{subcaption}

\usepackage{hyperref}
\hypersetup{colorlinks, linkcolor=Magenta,citecolor=Cerulean
}

\newcommand{\be}{\begin{equation}}
	\newcommand{\ee}{\end{equation}}
\newcommand{\bea}{\begin{eqnarray}}
	\newcommand{\eea}{\end{eqnarray}}
\newcommand{\vv}{``}

\begin{document}
\graphicspath{{FIGURE/}}
\topmargin=-1cm
	
\begin{center} 
{\large
{\bf 
Diffeomorphism invariance of the effective gravitational action}
}\\

\vspace*{0.6 cm}
		
C. Branchina\label{one}$^{\,a}$,
V. Branchina\label{two}$^{\,b}$, 
F.
Contino\label{three}$^{\,c}$,
R. Gandolfo\label{four}$^{\,b}$,
A.
Pernace\label{five}$^{\,b,\,d}$
		\vspace*{0.1cm}

\vskip12pt	

{\it			
			${}^a${\footnotesize Department of Physics, University of Calabria, and INFN-Cosenza,
			Arcavacata di Rende, I-87036, Cosenza, Italy}
			
			\vskip 5pt
			
			${}^b${\footnotesize Department of Physics, University of Catania, and INFN-Catania,
			Via Santa Sofia 64, I-95123 
			Catania, Italy}
			
			\vskip 5pt
			
			${}^c${\footnotesize Scuola Superiore Meridionale, Largo San Marcellino 10, 80138 Napoli, Italy}
			
			\vskip 5pt
			
			${}^d${\footnotesize Centro de Física Teórica e Computacional, Faculdade de Ciências, Universidade de Lisboa}
		}
			
	 \vskip 20pt
	 {\bf Abstract}
	 \noindent
		 
\end{center}

{\small 
	
\noindent
We  investigate on the diffeomorphism invariance of the effective gravitational action, focusing in particular on the path integral measure. In the literature, two different measures are mainly considered, the Fradkin\,-Vilkovisky and the Fujikawa one.  With the help of  detailed calculations, 
we show that, despite claims to the contrary, the Fradkin\,-Vilkovisky measure is diffeomorphism invariant, while the Fujikawa measure is not.
In particular, we see that, contrary to na\"ive expectations, the presence of  $g^{00}$ factors in the Fradkin\,-Vilkovisky measure is necessary to ensure the invariance of the effective gravitational action. We also comment on results recently appeared in the literature, and show that formal calculations can easily miss delicate points. }

\section{Introduction}
\label{Intro}
\numberwithin{equation}{section}

If quantum gravity is described as a quantum field theory (as opposed to string theory or other formulations), two cases are possible. It might be either a UV-complete theory or an effective field theory (EFT) \cite{Weinberg:1996kw}. 
In each of these cases, it is defined through the path integral
\begin{equation}\label{pathintegral}
	\int {\rm d}\mu \,e^{-S}\,,
\end{equation}
where ${\rm d}\mu$ is the measure and $S$ the classical action. Of course, only after the theory $({\rm d}\mu,S)$ is given can the possible existence of fixed points be investigated\footnote{Needless to say, if ${\rm d}\mu'$ and $S'$  are such that  ${\rm d}\mu' \,e^{-S'}= {\rm d}\mu \,e^{-S}$, the couple $({\rm d}\mu',S')$ defines the same theory as  $({\rm d}\mu,S)$. Obviously, the RG properties of the theory (including the presence/absence of fixed points) are not altered by such a trivial reshuffling of terms.}. In particular, it can be seen whether the theory is UV-complete, in which case it possesses a UV-attractive fixed point, or it is an EFT valid up to a certain energy scale, above which its UV-completion takes over. Starting from the path integral\,\eqref{pathintegral}, the Wilsonian RG approach implements a non-perturbative  definition of the theory (successive elimination of infinitesimal shells of modes), no matter if the theory is UV-complete or an EFT.

In a recent paper \cite{Branchina:2024xzh}, we calculate the one-loop (euclidean) effective action $\Gamma^{1l}$ in quantum gravity within the Einstein-Hilbert truncation, paying due attention to the path integral measure and to the identification of the physical UV cutoff $\Lambda$. In \cite{Branchina:2024lai}, we extend this analysis deriving the renormalization group (RG) equations for the running Newton and cosmological constant. In \cite{Branchina:2024xzh}, we show that the quantum correction to the vacuum energy does not present 
quartically and quadratically divergent contributions\footnote{For other related approaches to the naturalness problem see \cite{Branchina:2022gll,Branchina:2022jqc, Branchina:2023ogv, Branchina:2023rgi, Branchina:2024ljd}.}.
In \cite{Branchina:2024lai}, we find that the RG flow of the Newton and cosmological constant significantly differs from the so-called asymptotic safety (AS) scenario \cite{Reuter:1996cp,Souma:1999at,Reuter:2001ag,Bonanno:2004sy}. In particular, we do not see any sign of a non-trivial UV-attractive fixed point, which is the distinguishing feature of this latter scenario.

In \cite{Branchina:2024xzh,Branchina:2024lai}, the Fradkin\,-Vilkovisky (FV) path integral measure \cite{Fradkin:1973wke, Fradkin:1975sj,Fradkin:1977hw} is used. It is sometimes claimed that, since this measure contains non-covariant factors of the time-time component $g^{00}(x)$ of the inverse metric, it is not diffeomorphism invariant, and that (on the contrary) the invariant measure is the one proposed by Fujikawa \cite{Fujikawa:1983im}. This point of view has been recently taken up in \cite{Bonanno:2025xdg} where, considering the case of a scalar theory in a gravitational background, the authors claim to demonstrate that the Fujikawa measure is diffeomorphism invariant (they then extend  these conclusions to the case of pure gravity).

The aim of the present paper is to bring clarity on the question related to the measure. Performing a thorough investigation, we show that, despite the presence of $g^{00}$ factors, the FV measure is diffeomorphism invariant. With the help of detailed calculations, we will show that this result is intimately related to the {\it very definition} of the path integral, that involves the introduction of a time ordering parameter and of a discretization of spacetime (lattice). By the same token, we will see that the result by which the Fujikawa measure appears to be diffeomorphism invariant is due to a {\it formal} treatment of the path integral, resorting to which the aforementioned points are overlooked. As a consequence, non-trivial terms that appear when the path integral undergoes a general coordinate transformation are missed. Once these terms are taken into account, it turns out that the FV measure is diffeomorphism invariant, while the Fujikawa measure is not.

The rest of the paper is organized as follows. In section \ref{diffinvmeas}, we consider the contribution to the effective gravitational action $\Gamma[g]$ due to the quantum fluctuations of a scalar field and start the investigation on its diffeomorphism invariance. Section \ref{transfcoeff} is devoted to the detailed calculation of all the terms that appear when $\Gamma[g]$ undergoes a general coordinate transformation. With the help of the results of section \ref{transfcoeff}, in section \ref{FVandFuji} we show that the FV measure is invariant, while the Fujikawa one is not. In section \ref{comments}, we compare our results with those of previous literature, in particular with those of the recent paper \cite{Bonanno:2025xdg}. Section \ref{Conclusions} is for the conclusions.

\section{Diffeomorphisms and path integral}
\label{diffinvmeas}

Let us consider the gravitational action $S_{\rm g}[g]$ (think for instance of the Einstein-Hilbert action) in the presence of a scalar field with action $S_{\rm m}[\phi,g]$. The total action is (from now on we indicate with $g$ the metric $g_{\mu\nu}$ when it appears as the argument of a function/functional, its determinant otherwise; the signature {\footnotesize $(-,+,+,+)$} is used)
\begin{align}\label{actionx}
	S[\phi,g]&=S_{\rm g}[g]+S_{\rm m}[\phi,g]=S_{\rm g}[g]+\int \dd[4]{x} \mathcal{L}_{\rm m}(\phi(x),\partial_\mu\phi(x),g(x))
\end{align} 
where $\mathcal{L}_{\rm m}(\phi(x),\partial_\mu\phi(x),g(x))$ is the matter Lagrangian density
\begin{equation}\label{matterlagr}
	\mathcal{L}_{\rm m}(\phi(x),\partial_\mu\phi(x),g(x))=-\frac12\sqrt{-g(x)} \left( g^{\mu\nu}(x) \partial_\mu \phi(x) \partial_\nu \phi(x) + m^2 \phi^2(x) \right)\,.
\end{equation}
The effective gravitational action $\Gamma[g]$ is given by
\begin{equation}\label{Zx}
	e^{i\Gamma[g]}=\int \,e^{i S[\phi(x),\,g(x)]}\prod_{ x}\Big[M(g( x))\dd{\phi( x)}\Big]\,,
\end{equation}
where $M(g(x))$ is a non-trivial term in the configuration space measure {\small $\prod_{x}\big[M(g( x))\dd{\phi( x)}\big]$}. In the literature, two different expressions for $M(g( x))$ are mainly considered. For the FV measure \cite{Fradkin:1973wke}, 
\begin{equation}\label{FVmeas}
	M_{_{\rm FV}}(g(x))=(-g^{00}(x))^{1/2}(-g(x))^{1/4}\,,
\end{equation}
while for the Fujikawa measure \cite{Fujikawa:1983im} ($\mu$ is an arbitrary mass scale)
\begin{equation}\label{Fujimeas}
	M_{_{\rm Fuji}}(g(x))=\mu\,(-g(x))^{1/4}\,.
\end{equation}
Within the canonical formalism, the FV measure is obtained from the phase space path integral measure {\small $\prod_x\big[\dd{\pi(x)}\dd{\phi(x)}\big]$} (Liouville), while the Fujikawa measure \cite{Fujikawa:1983im} from {\small $\prod_x \big[(g^{00}(x))^{-1/2}\,\dd{\pi(x)}\dd{\phi(x)}\big]$} \cite{Toms:1986sh}. After integration over the conjugate momenta, the FV configuration space measure \eqref{FVmeas} contains $[g^{00}(x)]^{1/2}$ factors, while the Fujikawa measure \eqref{Fujimeas} does not. 

A fundamental aspect to be stressed is the following. 
In\,\eqref{Zx}, 
{\small $\prod_x$} indicates that in the definition of the path integral a discretization of spacetime is introduced, which in turn implies that {\small $\prod_{ x}\big[M(g( x))\dd{\phi( x)}\big]$} in\,\eqref{Zx} can (obviously) be written as {\small $\big[\prod_{ x}M(g( x))\big]\big[\prod_{ x}\dd{\phi(x)}\big]$}. This latter (trivial) observation will be useful in the following. Moreover, since in section \ref{comments} we will  compare our results with those of \cite{Bonanno:2025xdg}, it is useful to observe that, at the beginning of their work, the authors define the path integral measure in the same way as we do in\,\eqref{Zx} (see their equations \,(2.1) and\,(2.2)). However, they are not entirely coherent with this definition along their calculations, and thus end up missing important terms in the transformation of the effective action under diffeomorphisms. We will add further comments on these points in due time.  

Let us go back now to the meaning of $\prod_x$, stressing a crucial  point first raised in \cite{Leutwyler:1964wn}, and later deeply investigated in \cite{Fradkin:1973wke}. In any formulation of 
quantum field theory, and in particular in the formulation of gauge theories, the construction of the basic transition amplitudes $\braket{\phi',t'}{\phi'',t''}$ (definition of the $S$ matrix) requires the identification of a parameter in terms of which a time ordering is introduced. Differently from the case of Yang-Mills theories, where the gauge transformation only affects the form of the fields, in gravity a general coordinate transformation acts on the coordinates, and thus on the argument of the fields. As we will see, this has a non-trivial impact on the time ordering. In the path integral formulation, the time ordering parameter can be introduced as follows. Given a coordinate system $x^{\mu}$, if the points of the lattice are obtained from the intersection between hypersurfaces $x^0={\rm const}$ and curves $x^i={\rm const}$, the time ordering parameter is identified\footnote{\label{sphyp}The quantization can also be realized considering a more general space-like hypersurface $\tau(x)={\rm const}$, in which case the role of ordering parameter is played by $\tau$.} with $x^0$ \cite{Leutwyler:1964wn,Fradkin:1973wke}. The notation $\prod_x$ in Eq.\,\eqref{Zx} indicates the product over all the points $Q_i$ of the lattice $\mathcal{E}_1$ defined in this way. 
Note that the construction of the lattice and the identification of the ordering parameter can be done only once the coordinate system has been specified. 

Let us call $\Sigma$ the reference frame with coordinates $x$ and $\hat \Sigma$ another frame with coordinates $\hat x$. In $\hat \Sigma$, the action\,\eqref{actionx} is written as (we use the notation $\text{\footnotesize $\hat \partial_\mu\equiv$}\pdv{}{\hat x^\mu}$)
\begin{equation}\label{actionhatx}
	\hat S[\hat\phi,\hat g]=\hat S_{\rm g}[\hat g]+\hat S_{\rm m}[\hat\phi,\hat g]=\hat S_{\rm g}[\hat g]+\int \dd[4]{\hat x} \mathcal{L}_{\rm m}(\hat \phi(\hat x),\hat\partial_\mu\hat\phi(\hat x),\hat g(\hat x))\,,
\end{equation}
and for the effective action $\hat \Gamma[\hat g]$ we have
\begin{equation}\label{Zhatx}
	e^{i\hat \Gamma [\hat g]}=\int \,e^{i\,\hat S[\hat\phi,\hat g]}\prod_{\hat x}\Big[M(\hat g(\hat x))\dd{\hat\phi(\hat x)}\Big]\,.
\end{equation}
Similarly to what we said for\,\eqref{Zx}, the notation $\prod_{\hat x}$ in\,\eqref{Zhatx} indicates the product over the points $P_i$ of the lattice $\mathcal{E}_2$ obtained from the intersection of hypersurfaces $\hat x^0={\rm const}$ with curves $\hat x^i={\rm const}$. In this case, the time ordering parameter is $\hat x^0$. We will see that, due to this change in the time ordering parameter, the path integral measure transforms non-trivially under diffeomorphisms\footnote{With an abuse of notation, throughout the paper we will use the same symbol $\hat x$ to indicate both the action of the diffeomorphism on $x$ and its corresponding image, i.e.\,\,$\hat x=\hat x(x)$.}. This important aspect is missed in \cite{Bonanno:2025xdg}: the measure {\small $\prod_{x}\big[M(g( x))\dd{\phi( x)}\big]$} should not be treated formally, and attention should be paid to the discretization and to the time ordering underlying this expression.

Let us investigate now  on the diffeomorphism invariance of the effective action, performing the transition from $\hat \Sigma$ to $\Sigma$. Our aim is to see for which choice of $M(g(x))$, either the FV one in\,\eqref{FVmeas} or the Fujikawa one in\,\eqref{Fujimeas}, we have $\hat \Gamma[\hat g]=\Gamma[g]$. Since the classical action is diffeomorphism invariant, i.e.\,$\hat S[\hat\phi,\hat g]=S[\phi,g]$, from\,\eqref{Zx} and\,\eqref{Zhatx} we see that $\hat \Gamma[\hat g]=\Gamma[g]$ if
\begin{equation}\label{measinv}
	\prod_{\hat x}\Big[M(\hat g(\hat x))\dd{\hat\phi(\hat x)}\Big]=\prod_{ x}\Big[M(g( x))\dd{\phi( x)}\Big]\,.
\end{equation} 
We stress again that in
$\hat \Sigma$ and $\Sigma$ two different lattices and time ordering parameters are considered; this is encoded in the two product symbols $\prod_{\hat x}$ of\,\eqref{Zhatx} and $\prod_{x}$ of\,\eqref{Zx}. To realize the bridge between $\hat \Gamma[\hat g]$ (defined in $\hat \Sigma$) and $\Gamma[g]$ (defined in $\Sigma$), we now proceed in two steps.

In the first step, while remaining in the frame $\hat \Sigma$, we move from the lattice $\mathcal E_2$ of points $P_i$ (and time ordering parameter $\hat x^0$) to the lattice $\mathcal E_1$ of points $Q_i$ (and time ordering parameter $x^0$). This is realized going from $\prod_{\hat x}\big[M(\hat g(\hat x))\dd{\hat\phi(\hat x)}\big]$ to $\prod_{x}\big[M(\hat g(\hat x(x)))\dd{\hat\phi(\hat x(x))}\big]$ and performing in the action $\hat S[\hat\phi,\hat g]$ the change of integration variables from $\hat x$ to $x$ through the relations $\hat x=\hat x(x)$, $\hat \phi(\hat x)=\hat \phi(\hat x(x))$ and $\hat g_{\mu\nu}(\hat x)=\hat g_{\mu\nu}(\hat x(x))$. We get (see comments  below\,\eqref{acx2} for the factors $A$ and $B$)
{\small\begin{equation}\label{Zhatx2}
	e^{i\hat \Gamma [\hat g]}=\int \,e^{i\,\hat S[\hat\phi,\hat g]}\Big[\prod_{\hat x} M(\hat g(\hat x))\Big]\Big[\prod_{\hat x}\dd{\hat\phi(\hat x)}\Big]=\int \,e^{i\,\hat S[\hat\phi,\hat g]}\Big[B\prod_{x} M(\hat g(\hat x(x)))\Big]\Big[A\prod_{x}\dd{\hat\phi(\hat x(x))}\Big]\,,
\end{equation}}

\noindent
where $\hat S$ in the right-hand side is written as
\begin{align}\label{acx2}
	&\hat S[\hat\phi,\hat g]=\hat S_{\rm g}[\hat g]+\hat S_{\rm m}[\hat\phi,\hat g]\nonumber\\
	&=\hat S_{\rm g}[\hat g]-\frac{1}{2} \int \dd[4]{x} J \sqrt{-\hat g(\hat x(x))} \left( \hat g^{\mu\nu}(\hat x(x))\pdv{x^{\rho}}{\hat x^{\mu}}\pdv{x^{\sigma}}{\hat x^{\nu}} \partial_\rho \hat\phi(\hat x(x)) \partial_\sigma \hat\phi(\hat x(x)) + m^2 \hat\phi^2(\hat x(x)) \right)\nonumber\\
	&\equiv\hat S_{\rm g}[\hat g]+\int \dd[4]{x} \widetilde{\mathcal{L}}_{\rm m}(\hat \phi(\hat x(x)),\partial_\mu\hat\phi(\hat x(x)),\hat g(\hat x(x)))\,,
\end{align}

\noindent
with $J\equiv \big|{\rm det}\,\pdv{\hat x}{x}\,\big|$ the Jacobian. Note that $\hat S[\hat\phi,\hat g]$ is the action for the fields $\hat \phi$ and $\hat g_{\mu\nu}$, that is the action in the frame $\hat \Sigma$. We stress that up to now only a change of integration variables has been performed (not a change of reference frame).
Moreover, $\widetilde{\mathcal L}_{\rm m}$ in the third line of \eqref{acx2} is the matter Lagrangian density\footnote{It is easy to see that $\widetilde{\mathcal L}_{\rm m}$ gives rise to the same equations of motion and the same energy-momentum tensor for the field $\hat \phi$ as $\mathcal{L}_{\rm m}$ in\,\eqref{actionhatx}. The only difference is that they are written in terms of the variables $x$ using the relations $\hat x=\hat x(x)$, $\hat \phi(\hat x)=\hat \phi(\hat x(x))$ and $\hat g_{\mu\nu}(\hat x)=\hat g_{\mu\nu}(\hat x(x))$.}, still in the frame $\hat \Sigma$, written in terms of the variables $x$. 

Let us comment now on the  factors $A$ and $B$ that appear in the right hand side of\,\eqref{Zhatx2}. One might na\"ively expect that the step performed above ($\prod_{\hat x}\to\prod_x$) consists of a trivial reshuffling of (spacetime) points, in which case one would have $A=B=1$. This is what is implicitly implemented in \cite{Bonanno:2025xdg}. However, we have seen that in going from the left to the right hand side of \eqref{Zhatx2}, a non-trivial change of lattice and of ordering parameter occurs (see above). This is what ultimately leads to the appearance of non-trivial factors $A\neq1$ and $B\neq1$ in\,\eqref{Zhatx2}. The latter are calculated in the next section.

In the second step, we move from $\hat \phi(\hat x(x))$  and $\hat g_{\mu\nu}(\hat x(x))$ to $\phi(x)$ and $g_{\mu\nu}(x)$. This will  eventually lead to the relation between $\hat \Gamma[\hat g]$ and $\Gamma[g]$. From\,\eqref{Zhatx2}, we get (below the invariance of the classical action, $\hat S[\hat\phi,\hat g]=S[\phi,g]$, is used)
{\footnotesize \begin{align}
	e^{i\hat \Gamma [\hat g]}&=\int \,e^{i\,\hat S[\hat\phi,\hat g]}\Big[B\prod_{x} M(\hat g(\hat x(x)))\Big]\Big[A\prod_{x}\dd{\hat\phi(\hat x(x))}\Big]=\int \,e^{i\,S[\phi, g]}\Big[B\, C\prod_{x} M(g(x))\Big]\Big[A\, E\prod_{x}\dd{\phi(x)}\Big]\label{Zhatxfinal}\,.
\end{align}}

\noindent
In the above equation,  the factor $C$ arises when we express $M(\hat g(\hat x(x)))$ in terms of $g_{\mu\nu}(x)$, while $E$ is the Jacobian of the transformation $\hat \phi(\hat x(x))\to\phi(x)$. As for $A$ and $B$, the calculation of $C$ and $E$ is performed in the next section.
 
From\,\eqref{Zx} and\,\eqref{Zhatxfinal},  we see that for the effective action to be invariant under general coordinate transformations, i.e. to have\,\,$\hat \Gamma [\hat g]=\Gamma [g]$, it is necessary that $A\,B\,C\,E=1$. In the coming sections \ref{transfcoeff} and \ref{FVandFuji}, we will see that while this is the case for the FV measure (see $M_{_{\rm FV}}$ in \eqref{FVmeas}), it fails to be true for the Fujikawa measure (see $M_{_{\rm Fuji}}$ in \eqref{Fujimeas}). Let us proceed now to the calculation of $A$, $B$, $C$ and $E$.

\section{Transformation factors}
\label{transfcoeff}

In the present section, we calculate the  transformation factors $A$, $B$, $C$ and $E$ that appear in \eqref{Zhatxfinal} following \cite{Fradkin:1973wke}, where similar calculations have been performed for the pure gravity case. In particular, we will find that the factors $A$ and $B$, that are related to the transition $\prod_{\hat x}\to\prod_x$ involved in the transformation of the effective action under diffeomorphisms, turn out to be non-trivial ($\neq1$). As anticipated, this shows that $\prod_{\hat x}\to\prod_x$ is not a trivial reshuffling of points, but rather a delicate step to be performed carefully while establishing how the effective action transforms under diffeomorphisms. Overlooking these terms, one is led to think that the FV measure is not diffeomorphism invariant, due to the presence of the $g^{00}$ factors (see\,\eqref{FVmeas}). In fact, the opposite is true. As it will be seen in section \ref{FVandFuji}, the invariance of the effective action emerges from a balance between these non-trivial terms and those coming from the $g^{00}$ factors in the FV measure. Let us proceed now with the calculation. 

\vskip 7pt
\noindent
\textbf{\textit{The factor E}} - We begin with the factor 
$E$. Since $\phi$ is a scalar field ({\small $\hat \phi(\hat x(x))=\phi(x)$}) and $E$ is the Jacobian of the transformation {\small $\hat \phi(\hat x(x))\to\phi(x)$}, clearly $E=1$. 

\vskip 7pt
\noindent
\textbf{\textit{The factor A}} -  
Let us move now to the factor $A$. 
As mentioned in the previous section, the FV measure {\small $\prod_x \big[M_{_{\rm FV}}(g(x))\dd{\phi(x)}\big]$} (see\,\eqref{FVmeas}) is obtained within the canonical formalism from the phase space path integral measure {\small $\prod_x\big[\dd{\pi(x)}\dd{\phi(x)}\big]$}, while the Fujikawa measure {\small $\prod_x \big[M_{_{\rm Fuji}}(g(x))\dd{\phi(x)}\big]$} (see\,\eqref{Fujimeas}) from {\footnotesize $\prod_x \big[(g^{00}(x))^{-1/2}\,\dd{\pi(x)}\dd{\phi(x)}\big]$}. To calculate $A$, we consider both cases at once writing (below {\small $f(g)=1$} for the FV measure and {\small $f(g)=(g^{00}(x))^{-1/2}$} for the Fujikawa one)
\begin{equation}\label{Gamma5}
	e^{i\Gamma_{\text{\tiny$\mathcal E_1$}}[g]}=e^{i S_{\rm g}[g]}\int \,e^{i\int\dd[4]{x}\left[\partial_0\phi (x)\pi( x)-\mathcal H(\pi(x),\,\phi(x))\right]}\Big[\prod_{x}f(g(x))\,\dd{\pi( x)}\dd{\phi( x)}\Big]
\end{equation}
where $\mathcal H(\pi(x),\,\phi(x))$ is the Hamiltonian density (here we use the shorthand notation $\mathcal H(\pi,\phi)$ for the hamiltonian density $\mathcal H(\pi,\phi,\partial_i\phi,g)$), $\pi(x)$ the momentum conjugate to the field $\phi(x)$
\begin{equation}\label{conjmom1}
	\pi(x)=\pdv{\mathcal{L}_{\rm m}(\phi(x),\partial_\mu\phi(x),g(x))}{(\partial_0\phi(x))}\,,
\end{equation}
and $\mathcal{L}_{\rm m}(\phi(x),\partial_\mu\phi(x),g(x))$ given in\,\eqref{actionx} the Lagrangian density in the frame $\Sigma$. As stressed in the previous section (see comments below\,\eqref{Fujimeas}), $\prod_x$ in \eqref{Gamma5} means that the discretization is realized considering the points $Q_i$ of the lattice $\mathcal E_1$ and that the time ordering parameter is $x^0$ (this is further stressed by the subscript $\mathcal E_1$ in $\Gamma_{\text{\tiny$\mathcal E_1$}}[g]$). 
In this respect, we stress that $\pi(x)$ is correctly defined by\,\eqref{conjmom1} since in this case $x^0$ is the time ordering parameter. As said above (see footnote \ref{sphyp}), we could also quantize the theory considering a more general family of space-like hypersurfaces $\tau(x)={\rm const}$, in which case the time ordering parameter would be $\tau$, and the conjugate momenta would be obtained differentiating the Lagrangian with respect to $\partial_{\tau}\phi$. 
 
For the effective action in $\hat\Sigma$ we have (now the lattice $\mathcal{E}_2$ of points $P_i$ is used)
\begin{equation}\label{Gamma6}
	e^{i\hat\Gamma_{\text{\tiny$\mathcal E_2$}}[\hat g]}=e^{i \hat S_{\rm g}[\hat g]}\int \,e^{i\int\dd[4]{\hat x}\left[\hat \partial_0\hat \phi (\hat x)\hat \pi( \hat x)-\mathcal H(\hat \pi(\hat x),\,\hat \phi(\hat x))\right]}\Big[\prod_{\hat x}f(\hat g(\hat x))\,\dd{\hat \pi(\hat  x)}\dd{\hat \phi(\hat x)}\Big]
\end{equation}
where $\mathcal H(\hat\pi(\hat x),\,\hat\phi(\hat x))$ is the Hamiltonian density, $\hat\pi(\hat x)$ the momentum conjugate to the field $\hat\phi(\hat x)$
\begin{equation}\label{conjmom2}
	\hat\pi(\hat x)=\pdv{\mathcal{L}_{\rm m}(\hat \phi(\hat x),\hat\partial_\mu\hat\phi(\hat x),\hat g(\hat x))}{(\hat\partial_0\hat\phi(\hat x))}\,,
\end{equation}
and $\mathcal{L}_{\rm m}(\hat \phi(\hat x),\hat\partial_\mu\hat\phi(\hat x),\hat g(\hat x))$ the Lagrangian density in the frame $\hat\Sigma$, see\,\eqref{actionhatx}. The symbol $\prod_{\hat x}$ indicates that the discretization is realized considering the points $P_i$ of the lattice $\mathcal E_2$. The time ordering parameter is now $\hat x^0$, and accordingly the conjugate momentum $\hat\pi(\hat x)$ is given by\,\eqref{conjmom2}.

As explained in the previous section, the bridge between $\hat\Gamma[\hat g]$ and $\Gamma[g]$ is realized in two steps. In the first one (from which the factor $A$ emerges), while remaining in the frame $\hat \Sigma$ we write $\hat\Gamma[\hat g]$ switching from the lattice $\mathcal E_2$ to the lattice $\mathcal E_1$ (that in turn implies switching from the time ordering parameter $\hat x^0$ to $x^0$). Considering then $\mathcal E_1$, for $\hat\Gamma[\hat g]$ we have
{\small\begin{align}\label{Gamma7}
	&e^{i\hat\Gamma_{\text{\tiny$\mathcal E_1$}}[\hat g]}=e^{i \hat S_{\rm g}[\hat g]}\,\int \,e^{i\int\dd[4]{x}\left[ \partial_0\hat \phi (\hat x(x))\widetilde \pi( \hat x(x))-\widetilde{\mathcal H}(\widetilde \pi(\hat x(x)),\,\hat \phi(\hat x(x)))\right]}\text{\footnotesize$\Big[F\prod_{x}f(\hat g(\hat x(x)))\Big]\Big[\prod_{x}\,\dd{\widetilde \pi(\hat  x(x))}\dd{\hat \phi(\hat x(x))}\Big]$}\,.
\end{align}}

\noindent
In the above equation, $F$ is the factor that arises when $\prod_{\hat x}f(\hat g(\hat x))$ is written in terms of $\prod_x f(\hat g(\hat x(x)))$, i.e.\,\,when we switch from $\mathcal E_2$ to $\mathcal E_1$\,:\, {\small $\prod_{\hat x}f(\hat g(\hat x))=F\prod_{x}f(\hat g(\hat x(x)))$}. We will shortly see that we do not need to calculate explicitly the transformation factor $F$. Moreover, since in\,\eqref{Gamma7} the time ordering parameter is $x^0$, the Hamiltonian density to be used is {$\widetilde{\mathcal H}(\widetilde \pi(\hat x(x)),\,\hat \phi(\hat x(x)))$}, that corresponds to the Lagrangian density
{\small$\widetilde{\mathcal{L}}_{\rm m}(\hat \phi(\hat x(x)),\partial_\mu\hat\phi(\hat x(x)),\hat g(\hat x(x)))$} in \eqref{acx2}, and the momentum $\widetilde\pi(\hat x(x))$ conjugate to $\hat\phi(\hat x(x))$ is (see comments below\,\eqref{acx2} and\,\eqref{conjmom1})
\begin{equation}\label{conjmom3}
	\widetilde\pi(\hat x(x))=\pdv{\widetilde{\mathcal{L}}_{\rm m}(\hat \phi(\hat x(x)),\partial_\mu\hat\phi(\hat x(x)),\hat g(\hat x(x)))}{(\partial_0\hat\phi(\hat x(x)))}\,.
\end{equation}

Let us perform now in\,\eqref{Gamma6} the functional integration over $\hat\pi(\hat x)$. We get
\begin{equation}\label{Gamma8}
		e^{i\hat\Gamma_{\text{\tiny$\mathcal E_2$}}[\hat g]}=e^{i \hat S_{\rm g}[\hat g]}\Big[\prod_{\hat x}f(\hat g(\hat x))\Big]\int \,e^{i\hat S_{\rm m}[\hat\phi,\,\hat g]}\Big[\prod_{\hat x}W(\hat g(\hat x))\Big]\prod_{\hat x}\,\dd{\hat \phi(\hat x)}
\end{equation}
where we define
\begin{equation}\label{MP}
	W(\hat g(\hat x))\equiv(-\hat g(\hat x))^{1/4}(-\hat g^{00}(\hat x))^{1/2}\,.
\end{equation}
Similarly, the integration over $\widetilde\pi(\hat x(x))$ in\,\eqref{Gamma7} gives 
\begin{align}\label{Gamma9}
	&e^{i\hat\Gamma_{\text{\tiny$\mathcal E_1$}}[\hat g]}=e^{i \hat S_{\rm g}[\hat g]}\,F\,\Big[\prod_{x}f(\hat g(\hat x(x)))\Big]\,\int \,e^{i\hat S_{\rm m}[\hat\phi,\,\hat g]}\Big[\prod_{x}Y(\hat g(\hat x(x)))\Big]\prod_{x}\,\dd{\hat \phi(\hat x(x))}\,,
\end{align}
where we define ($J\equiv\Big|{\rm det}\,\pdv{\hat x}{x}\,\Big|$ is the Jacobian related to the change of variables $\hat x\to x$) 
\begin{equation}\label{MQ}
	Y(\hat g(\hat x(x)))\equiv J^{1/2}(-\hat g(\hat x(x)))^{1/4}\big[-\hat g^{\mu\nu}(\hat x(x))\,\hat\partial_\mu x^0\,\hat\partial_\nu x^0\,\big]^{1/2}\,.
\end{equation}
Moreover, writing $\prod_{\hat x}\dd{\hat\phi(\hat x)}$ in terms of $\prod_{x}\dd{\hat\phi(\hat x(x))}$ (see Eq.\,\eqref{Zhatx2}),  Eq.\,\eqref{Gamma8} can be written as 
\begin{equation}\label{Gamma10}
	e^{i\hat\Gamma_{\text{\tiny$\mathcal E_2$}}[\hat g]}=e^{i \hat S_{\rm g}[\hat g]}\Big[\prod_{\hat x}f(\hat g(\hat x))\Big]\int \,e^{i\hat S_{\rm m}[\hat\phi,\,\hat g]}\Big[\prod_{\hat x}W(\hat g(\hat x))\Big]\,\Big[A\prod_{x}\,\dd{\hat \phi(\hat x(x))}\Big]\,.
\end{equation}
Since $\hat\Gamma_{\text{\tiny$\mathcal E_2$}}[\hat g]$ and $\hat\Gamma_{\text{\tiny$\mathcal E_1$}}[\hat g]$ are the same effective action in $\hat \Sigma$, simply written using the two lattices $\mathcal E_2$ and $\mathcal E_1$, respectively, from\,\eqref{Gamma9} and\,\eqref{Gamma10} we finally get 
\begin{equation}\label{J1}
	A=\frac{\prod_{x}Y(\hat g(\hat x(x)))}{\prod_{\hat x}W(\hat g(\hat x))}\,.
\end{equation}

We now write $A$ in a convenient form. Let us consider an infinitesimal coordinate transformation (in the following we only need to keep terms up to first order in $\varepsilon$)
\begin{equation}\label{coordtransf}
	\hat x^\mu=x^\mu+\varepsilon^\mu(x)\,,
\end{equation}
and switch from $\prod_x$ to $\prod_{\hat x}$ in the numerator of \eqref{J1}. As we will see below (Eq.\,\eqref{interm}), this allows to factor out a term that cancels the denominator. Observing that (as usual, {\small $\delta^{(4)}(0)$} comes from $\sum_x\to\int\text{\small$\dd[4]{x}$}$)
\begin{align}\label{A}
	\prod_{x} J^{1/2}=\exp(\frac{\delta^{(4)}(0)}{2}\int\dd[4]{x}\log (1+\partial_\mu \varepsilon^\mu))=\exp(\frac{\delta^{(4)}(0)}{2}\int\dd[4]{x}\partial_\mu \varepsilon^\mu)=1\,,
\end{align}
where we have used
\begin{equation}\label{J}
	\pdv{\hat x^\mu}{x^\nu}=\delta^\mu_\nu+\partial_\nu\varepsilon^\mu\quad;\quad \dd[4]{\hat x}=\dd[4]{x}J\equiv\dd[4]{x}\Big|{\rm det}\,\pdv{\hat x}{x}\,\Big|=\dd[4]{x}(1+\partial_\mu\varepsilon^\mu(x))\,,
\end{equation}
and calculating the third factor in the right hand side of \eqref{MQ} (\,$\hat\partial_\mu x^0=\delta_\mu^0-\partial_\mu\varepsilon^0$\,),
\begin{align}\label{calculations}
	-\hat g^{\mu\nu}(\hat x(x))\,\hat\partial_\mu x^0\,\hat\partial_\nu x^0=-\hat g^{00}+2\,\hat g^{0\mu}\,\partial_\mu \varepsilon^0=-\hat g^{00}\Big(1-2\,\frac{\hat g^{0\mu}}{\hat g^{00}}\,\partial_\mu \varepsilon^0\Big)\,,
\end{align}
from\,\eqref{MP},\,\eqref{MQ},\,\eqref{A},\,\eqref{calculations} we get
{\small\begin{align}
	&\prod_{x}Y(\hat g(\hat x(x)))=\exp(\frac{\delta^{(4)}(0)}{2}\int\dd[4]{\hat x}(1-\partial_\mu\varepsilon^\mu)\log\Big[(-\hat g(\hat x))^{1/2}\big(-\hat g^{00}(\hat x)\big)\Big(1-2\,\frac{\hat g^{0\mu}(\hat x)}{\hat g^{00}(\hat x)}\,\partial_\mu \varepsilon^0\,\Big)\Big])\nonumber\\
	&=\Big[\prod_{\hat x}W(\hat g(\hat x))\Big]\,\exp(-\frac{\delta^{(4)}(0)}{2}\int\dd[4]{\hat x}\Big[\partial_\mu\varepsilon^\mu\log\Big((-\hat g(\hat x))^{1/2}\big(-\hat g^{00}(\hat x)\big)\Big)+2\,\frac{\hat g^{0\mu}(\hat x)}{\hat g^{00}(\hat x)}\,\partial_\mu \varepsilon^0\Big]\,)\label{interm}\,.
\end{align}}

\noindent
Finally, inserting\,\eqref{interm} in\,\eqref{J1} we have
\begin{equation}\label{A1}
	A=\exp(-\frac{\delta^{(4)}(0)}{2}\int\dd[4]{x}\Big[\partial_\mu\varepsilon^\mu\log\Big((-g(x))^{1/2}\big(-g^{00}(x)\big)\Big)+2\,\frac{g^{0\mu}(x)}{g^{00}(x)}\,\partial_\mu \varepsilon^0\Big]\,)\,.
\end{equation}

\vskip 5pt
\noindent
\textbf{\textit{The factor B}} -
Let us calculate now $B$. We begin by observing that $M(\hat g(\hat x))$ in\,\,\eqref{Zhatx2} is nothing but the product $f(\hat g(\hat x))W(\hat g(\hat x))$ in\,\,\eqref{Gamma8}. We then have (below we make use of\,\eqref{J})
\begin{align}
	&\prod_{\hat x} M(\hat g(\hat x))=\prod_{\hat x}\big[f(\hat g(\hat x))W(\hat g(\hat x))\big]=\exp(\delta^{(4)}(0)\int\dd[4]{\hat x}\log(f(\hat g(\hat x))W(\hat g(\hat x))))\nonumber\\
	&=\exp(\delta^{(4)}(0)\int\dd[4]{x}\partial_\mu\varepsilon^\mu\log(f(g(x))W(g(x))))\prod_{x}\big[f(\hat g(\hat x(x)))W(\hat g(\hat x(x)))\big]\nonumber\\
	&=\exp(\delta^{(4)}(0)\int\dd[4]{x}\partial_\mu\varepsilon^\mu\log(M(g(x))))\prod_{x} M(\hat g(\hat x(x)))\label{B1}\,.
\end{align}
Recalling that $\prod_{\hat x} M(\hat g(\hat x))=B\prod_{x} M(\hat g(\hat x(x)))$ (see\,\eqref{Zhatx2}), from \eqref{B1} we have 
\begin{equation}\label{B}
	B=\exp(\delta^{(4)}(0)\int\dd[4]{x}\partial_\mu\varepsilon^\mu\log(M(g(x))))\,.
\end{equation}

\vskip 5pt
\noindent
\textbf{\textit{The factor C}} \,-\,
Finally, we calculate the factor $C$, that is defined by $\prod_x M(\hat g(\hat x(x)))=C \prod_x M(g(x))$ (see\,\eqref{Zhatxfinal}). At first order in $\varepsilon$, we have
\begin{align}
	&M(\hat g^{\mu\nu}(\hat x(x)))=M\big(g^{\mu\nu}(x)+g^{\mu\sigma}(x)\partial_\sigma\varepsilon^\nu(x)+g^{\nu\sigma}(x)\partial_\sigma\varepsilon^\mu(x)\big)=\nonumber\\
	&=M(g)+\pdv{M(g)}{g^{\alpha\beta}(x)}\, (g^{\alpha\sigma}(x)\partial_\sigma\varepsilon^\beta(x)+g^{\beta\sigma}(x)\partial_\sigma\varepsilon^\alpha(x))\,,
\end{align}
where we used
\begin{equation}\label{metric}
	\hat g^{\mu\nu}(\hat x)=\pdv{\hat x^{\mu}(x)}{x^\rho}\pdv{\hat x^{\nu}(x)}{x^\sigma}\,g^{\rho\sigma}(x)=g^{\mu\nu}(x)+g^{\mu\sigma}(x)\partial_\sigma\varepsilon^\nu(x)+g^{\nu\sigma}(x)\partial_\sigma\varepsilon^\mu(x)\,.
\end{equation}
We then get
\begin{align}
	&\prod_x M(\hat g(\hat x(x)))=\exp(\delta^{(4)}(0)\int\dd[4]{x}\log\Big[M(g)+\pdv{M(g)}{g^{\alpha\beta}(x)}\, (g^{\alpha\sigma}(x)\partial_\sigma\varepsilon^\beta(x)+g^{\beta\sigma}(x)\partial_\sigma\varepsilon^\alpha(x))\Big])\nonumber\\
	&=\exp(\delta^{(4)}(0)\int\dd[4]{x}[M(g)]^{-1}\pdv{M(g)}{g^{\alpha\beta}(x)}\, (g^{\alpha\sigma}(x)\partial_\sigma\varepsilon^\beta(x)+g^{\beta\sigma}(x)\partial_\sigma\varepsilon^\alpha(x)))\prod_x M(g(x))\,,
\end{align}
from which
\begin{equation}\label{C}
	C=\exp(\delta^{(4)}(0)\int\dd[4]{x}[M(g)]^{-1}\pdv{M(g)}{g^{\alpha\beta}(x)}\, (g^{\alpha\sigma}(x)\partial_\sigma\varepsilon^\beta(x)+g^{\beta\sigma}(x)\partial_\sigma\varepsilon^\alpha(x)))\,.
\end{equation}

Having calculated the factors $A$, $B$, $C$ and $E$, we can now see how the FV and the Fujikawa measure transform under diffeomorphisms. This is the subject of the next section.

\section{Fradkin\,-Vilkovisky versus Fujikawa measure}
\label{FVandFuji}
As shown in section \ref{diffinvmeas} (see\,\eqref{Zhatxfinal} and comments below), for the effective action to be diffeomorphism invariant, it must be $A\,B\,C\,E=1$. Having found in the previous section the general expressions for $A$, $B$, $C$ and\,$E$, we can now calculate these terms for both $M(g(x))=M_{_{\rm FV}}(g(x))$ and $M(g(x))=M_{_{\rm Fuji}}(g(x))$.
From a simple inspection of\,\eqref{A1},\,\eqref{B} and\,\eqref{C} we see that, while $B$ and $C$ depend on the specific form of $M(g(x))$, $A$ does not. Moreover, we have already seen at the beginning of section\,\ref{transfcoeff} that $E=1$. 
We have then to calculate $B$ and $C$ for $M_{_{\rm FV}}(g(x))$ and $M_{_{\rm Fuji}}(g(x))$, respectively.

\vskip 7pt
\noindent
\textbf{\textit{Fradkin\,-Vilkovisky measure}} - Inserting\,\eqref{FVmeas} in\,\eqref{B} and\,\eqref{C} we get
\begin{align}
	B_{_{\rm FV}}&=\exp(\frac{\delta^{(4)}(0)}{2}\int\dd[4]{x} \partial_\mu\varepsilon^\mu\log[(-g(x))^{1/2}(-g^{00}(x))])\label{BFV}\\
	C_{_{\rm FV}}&=\exp(\delta^{(4)}(0)\int\dd[4]{x} \frac{g^{0\mu}(x)}{g^{00}(x)}\,\partial_\mu\varepsilon^0)\label{CFV}\,,
\end{align}
from which (see also\,\eqref{A1})
\begin{equation}\label{mira}
	B_{_{\rm FV}}\,C_{_{\rm FV}}=A^{-1}\,.
\end{equation}
Eq.\,\eqref{mira}, together with the result $E=1$, shows that for the FV measure
\begin{equation}\label{FV}
	A\,B_{_{\rm FV}}\, C_{_{\rm FV}}\,E=1\,,
\end{equation}
which means that in this case (see\,\eqref{Zhatxfinal}):\, $\hat \Gamma [\hat g]=\Gamma [g]$.
As anticipated, the effective action calculated using the FV measure {\it is} diffeomorphism invariant.

\vskip 7pt
\noindent
\textbf{\textit{Fujikawa measure}} - Inserting\,\eqref{Fujimeas} in\,\eqref{B} and\,\eqref{C} we get
\begin{align}
	B_{\rm Fuji}&=\exp(\delta^{(4)}(0)\int\dd[4]{x} \partial_\mu\varepsilon^\mu\log((-g(x))^{1/4}\mu))\label{BFuji}\\
	C_{\rm Fuji}&=1\label{CFuji}\,.
\end{align}
From\,\eqref{A1},\,\eqref{BFuji} and\,\eqref{CFuji} (together with $E=1$) we obtain
\begin{equation}\label{Fuji}
	A\,B_{_{\rm Fuji}}\, C_{_{\rm Fuji}}\,E=\exp(\delta^{(4)}(0)\int\dd[4]{x}\Big[ \partial_\mu\varepsilon^\mu\log(\frac{\mu}{(-g^{00})^{1/2}})-\frac{g^{0\mu}(x)}{g^{00}(x)}\,\partial_\mu\varepsilon^0\Big])\,,
\end{equation}
which means that in this case (see\,\eqref{Zhatxfinal}):\, $\hat \Gamma [\hat g]\neq\Gamma [g]$.
As anticipated, and contrary to what is claimed in \cite{Bonanno:2025xdg}, the effective action calculated using the Fujikawa measure {\it is not} diffeomorphism invariant.

\section{Comparison with existing literature}
\label{comments}

The main result of the analysis presented in the previous sections is that the FV measure \cite{Fradkin:1973wke} is diffeomorphism invariant, while the Fujikawa measure \cite{Fujikawa:1983im} is not. The na\"ive argument according to which the FV measure is claimed to be non-invariant is that it contains factors of the time-time component $g^{00}$ of the inverse metric. Actually, we have shown that the opposite is true. As thoroughly discussed in the previous sections, the $g^{00}$ factors in the FV measure guarantee the diffeomorphism invariance of the effective action. In fact, their presence is imposed by the necessity of having a time ordering parameter to define $S$ matrix elements. By the same token, we have shown that the Fujikawa measure is not diffeomorphism invariant. Moreover, our calculations allow to understand why in the recent work \cite{Bonanno:2025xdg} the opposite conclusion is reached, namely that the Fujikawa measure rather than the FV one is diffeomorphism invariant. The detailed analysis on the transformation of the effective action under diffeomorphisms developed in the previous sections shows that some of the terms involved in this transformation are missed in that paper.

Since the question of the diffeomorphism invariance of the effective action, sometimes source of discussions and controversies, is central to the present work, we find it worth  to make a more detailed comparison between our results and those of \cite{Bonanno:2025xdg}. We will see that the calculations and claims of that paper are flawed. 

The authors of \cite{Bonanno:2025xdg} consider the scalar theory defined by\,\eqref{actionx} and\,\eqref{matterlagr} above (see their section II). A simple inspection of their Eqs.\,(2.14) and\,(2.19) shows that they overlook the transition between the two lattices involved in the coordinate transformation (together with the related change of time ordering parameter). This is tantamount to assume that this transition is nothing but a trivial reshuffling of points (see\,\,\eqref{Zhatx2} and comments below). Actually, the authors of \cite{Bonanno:2025xdg} consider {\it only} the factors related to the transformation $(-\hat g)^{1/4}\hat\phi\to(-g)^{1/4}\phi$. According to the notation that we introduced in the previous sections, this amounts to consider {\it only} the factors $C$ and $E$ in\,\eqref{Zhatxfinal}, while missing both the non-trivial factors $A$ and $B$. It is precisely because they miss these latter terms that the authors of \cite{Bonanno:2025xdg} are led to conclude that the Fujikawa measure is diffeomorphism invariant. The same issues arise when the authors extend their considerations to pure quantum gravity (see their section VI). Also in this case, they consider the Fujikawa measure \cite{Fujikawa:1983im}, and attempt to demonstrate that the latter is BRST invariant. Once again, 
they miss non-trivial transformation factors in the transition $\hat x\to x$, as it is immediately seen in their Eqs.\,(6.3)-(6.10). 

Actually, all the arguments and conclusions in \cite{Bonanno:2025xdg} are flawed, since they are based on the alleged diffeomorphism invariance of the Fujikawa measure. For instance, they claim that the path integral measure to be used in phase space is \cite{Toms:1986sh} (see section III of \cite{Bonanno:2025xdg}, in particular the comments below Eqs.\,(3.9) and\,(3.27))
\begin{equation}\label{Toms}
	\prod_x \big[(g^{00}(x))^{-1/2}\,\dd{\pi(x)}\dd{\phi(x)}\big]\,.
\end{equation}
Starting with\,\eqref{Toms}, in fact, after integration over the conjugate momenta $\pi$ the Fujikawa measure in configuration space is obtained. As shown in the previous sections, however, the diffeomorphism invariant measure in configuration space is the Fradkin\,-Vilkovisky one. The latter is obtained after integration over $\pi$ starting from the phase space path integral measure
\begin{equation}\label{Liouville}
\prod_x \big[\dd{\pi(x)}\dd{\phi(x)}\big]\,,
\end{equation}
that is nothing but the Liouville measure. This is what one would expect from general considerations based on the path integral construction of the basic transition amplitude $\braket{\phi',t'}{\phi'',t''}$.

To support their claim on the diffeomorphism invariance of the Fujikawa measure, the authors of \cite{Bonanno:2025xdg} (see their section IV) consider the contribution $\delta S$ to the effective gravitational action from a free scalar
field, obtained using this measure ($\mu$ is an arbitrary mass scale in the Fujikawa measure, see\,\eqref{Fujimeas})
\begin{equation}\label{1leffacfuji}
	\delta S[g]=\frac12\Tr\log(\frac{-\square+m^2}{\mu^2})\,,
\end{equation}
where $-\square$ is the spin-$0$ Laplace-Beltrami operator for the metric $g_{\mu\nu}$. They evaluate the trace resorting to two different methods: (i) sum over the eigenvalues of\, $-\square+m^2$\,\,;\, (ii) proper-time formalism, and get for $\delta S$ the results\,(4.7) and\,(4.15) of \cite{Bonanno:2025xdg}, respectively. Their calculations, however, overlook a quite delicate point (see below). This oversight leads them in both cases to a result for $\delta S$ expressed in terms of diffeomorphism invariant quantities, so they conclude that the Fujikawa measure is diffeomorphism invariant.  The delicate point is that, in the case of gravitational theories, subtleties arise in the calculation of $\log(-\square+m^2)$ (that appears in their Eqs.\,(4.4) and\,(4.13)), and one should carefully take into account the distributional nature of the Green's function of $(-\square+m^2\,)$ \cite{Fradkin:1973wke, Fradkin:1976xa}. When this is done, a non-trivial term contributing to the trace in\,\eqref{1leffacfuji} is found. This term is missed in \cite{Bonanno:2025xdg}. Taking $x^0$ as time ordering parameter (see comments below Eqs.\,\eqref{Fujimeas} and\,\eqref{Zhatx}), this latter term turns out to be $\delta^{(4)}(0)\int\dd[4]{x}\log(g^{00})$ \cite{Fradkin:1976xa}, which is not diffeomorphism invariant.
As a consequence, the quantum correction $\delta S$ calculated with the Fujikawa measure {\it is not} diffeomorphism invariant. 

Let us consider now the result for $\delta S[g]$ obtained when the FV measure rather than the Fujikawa one is used, which is what we calculate in \cite{Branchina:2024xzh}. The result is similar to\,\eqref{1leffacfuji} (obtained with the Fujikawa measure), but in addition to the \vv\,$\Tr\log$\,'' it contains an extra term (see second line of\,\,\eqref{scalarres}), whose crucial importance will be soon clear. The calculation is performed taking the metric $g^{(a)}_{\mu\nu}$ of a sphere\footnote{This useful choice is often considered in the literature, see for instance \cite{Taylor:1989ua, Dou:1997fg, Becker:2021pwo, Ferrero:2024yvw}.} of radius $a$, and gives (see Eq.\,(48) of \cite{Branchina:2024xzh})
\begin{align}\label{scalarres}
	\delta S[g^{(a)}]&=\frac12\log\left[{\rm det}\left(-\widetilde\square+a^2 m^2\right)\right]-\frac12\log\Big(\prod_x \widetilde g^{\,00}(x)\Big)\nonumber\\
	&=\frac12\Tr\log\left(-\widetilde\square+a^2 m^2\right)-\frac{\delta^{(4)}(0)}{2}\int\dd[4]{x}\log(\widetilde g^{\,00}(x))\,,
\end{align}
where $-\widetilde\square$ is the Laplace-Beltrami operator for the metric $\widetilde g_{\mu\nu}$ of a unit sphere, and $\widetilde g^{\,00}$ is the time-time component of the inverse of $\widetilde g_{\mu\nu}$. To avoid misunderstandings, we stress again that, despite the presence in\,\eqref{scalarres} of $\widetilde g_{\mu\nu}$ and $-\widetilde\square$\,,\, $\delta S[g^{(a)}]$ is calculated for a sphere of {\it generic} radius $a$. As clearly explained in \cite{Branchina:2024xzh} and \cite{Branchina:2025kmd} (to which we refer for details), these dimensionless quantities ($-\widetilde\square$\,,\, $\widetilde g_{\mu\nu}$ and $a^2 m^2$) appear thanks to the presence of the FV measure. As we stressed several times in \cite{Branchina:2024xzh,Branchina:2024lai,Branchina:2025kmd}, and contrary to what is incorrectly claimed in \cite{Held:2025vkd}, we do not take the radius $a$ as reference scale. This is immediately evident if for instance we note that in\,\eqref{scalarres} the combination $a^2m^2$ appears.

Some crucial points need to be stressed. First of all, we observe that the term {\small $\frac12\log(\prod_x \widetilde g^{\,00}(x))$} in\,\eqref{scalarres} comes from the exponentiation of the measure term {\small $\prod_x (\widetilde g^{\,00}(x))^{1/2}$} (see\,\eqref{Zx}, with $M(g)$ given in\,\eqref{FVmeas}). In this respect, we recall (see comments below\,\eqref{Fujimeas}) that the discretization involved in the definition of the path integral, encoded in {\small $\prod_x$}, allows the trivial splitting {\small $\prod_{ x}\big[(g^{00}(x))^{1/2}(g(x))^{1/4}\dd{\phi( x)}\big]=\big[\prod_{x}(g^{00}(x))^{1/2}\big]\big[\prod_{ x}(g(x))^{1/4}\dd{\phi(x)}\big]$}, which is what we operate to get\,\eqref{scalarres}. Moreover, as it should be clear from the comments below\,\eqref{1leffacfuji}, the calculation of {\small $\frac12\Tr\log\small(-\widetilde\square+a^2 m^2\small)$} in\,\eqref{scalarres} gives rise to the term {\small $\frac{\delta^{(4)}(0)}{2}\int\dd[4]{x}\log(\widetilde g^{\,00}(x))$} \cite{Fradkin:1973wke, Fradkin:1976xa}. This term {\it cancels} the last one in the second line\footnote{See \cite{Leutwyler:1964wn} for earlier discussions on the appearance and cancellation of $\delta(0)$ divergences.} of\,\,\eqref{scalarres}. All the other terms that come from the \vv\,$\Tr\log$\,'' turn out to be diffeomorphism invariant. Therefore, the presence of the $g^{00}(x)$ factors in the measure not only does not spoil the diffeomorphism invariance of the effective action (as one would na\"ively expect) but it is rather necessary\footnote{In \cite{Branchina:2024xzh}, the term {\scriptsize $\frac12\log(\prod_x \widetilde g^{\,00}(x))=\frac{\delta^{(4)}(0)}{2}\int\dd[4]{x}\log(\widetilde g^{\,00}(x))$} is indicated with $\mathcal C$ (see Eq.\,(48) of \cite{Branchina:2024xzh}). There, we do not refer to the cancellation discussed in this work since the terms involved are $a$-independent, and have no impact in the calculation of the quantum corrections to the Newton and cosmological constant (the objective of \cite{Branchina:2024xzh}).} to {\it compensate} non-invariant terms that arise from the calculation of the \vv\,$\Tr\log$\,''. Similar considerations apply to the pure gravity case \cite{Fradkin:1973wke}.

We have just seen that $\delta S$ is given by\,\eqref{scalarres}, and that the presence of $g^{00}(x)$ factors in the configuration space measure is crucial to have a diffeomorphism invariant result. In \cite{Bonanno:2025xdg}, the more general measure {\small $\prod_{ x}\big[(g(x))^{1/4}\,\Omega_g(x)\dd{\phi( x)}\big]$} is also considered, and the authors write (see their section V)
\begin{equation}\label{1leffacgen}
	\delta S[g]=\frac12\Tr\log[\Omega_g^{-2}(-\square+m^2)]\,.
\end{equation} 
Based on their idea that the diffeomorphism invariance of $\delta S$ is ensured if the fluctuation operator $\Omega_g^{-2}(-\square+m^2)$ is covariant, they claim that the only possibility is for $\Omega_g$ is to be a scalar. The analysis of the present work (see in particular the three previous paragraphs) shows that, contrary to this claim, the only possibility to have a diffeomorphism invariant effective action is to take $\Omega_g=(g^{00})^{1/2}$. 

In fact, as explained in detail above, the diffeomorphism invariance of the effective action is not na\"ively related to the fact that only covariant functions and functionals are present in the path integral. The invariance emerges from a delicate balance between {\it all} the elements that enter the definition of the path integral. We have seen that the delicate point concerns the product\,\, $\prod_x$\, of all these elements, more specifically the time ordering and the discretization (lattice) involved in the definition of the path integral and encoded in this product.  

Going back to\,\eqref{1leffacgen}, for a scalar $\Omega_g$ the authors of \cite{Bonanno:2025xdg} manage to write $\Omega_g^{-2}(-\square+m^2)$ in the form of a minimal Laplace-type operator, i.e.\,\,an operator of the kind $(-\square+m^2\,)$ as the one in\,\eqref{1leffacfuji} (see Eqs.\,(A5) and\,(A6) of \cite{Bonanno:2025xdg}). They then calculate the \vv\,$\Tr\log$\,'' resorting to proper-time techniques, but fail to treat the $\log(-\square)$ with the due care, thus missing relevant terms (see the thorough discussion below\,\eqref{1leffacfuji})\cite{Fradkin:1976xa}. As a consequence, the proper-time RG equation\,(5.6) of \cite{Bonanno:2025xdg}, that the authors derive differentiating with respect to the scale $\Lambda$ their (incomplete) regularized action $S_{\Lambda}$, cannot be trusted, and the considerations and conclusions they draw starting from this equation are flawed. 

The authors then move to consider the pure gravity case (in the EH truncation; see section VII of \cite{Bonanno:2025xdg}), and develop considerations that mirror (as they say) those made for the contribution to the effective gravitational action $\Gamma[g]$ from the scalar field. Even in this pure gravity case, in fact, the contribution $\delta S_{\rm grav}$ to $\Gamma$ from the graviton is calculated considering traces similar to\,\eqref{1leffacgen}. They claim that if one uses the Fuijikawa measure \cite{Fujikawa:1983im} a diffeomorphism invariant RG flow is obtained that (in four dimensions) contains terms of the kind (see their Eqs.\,(7.1) and\,(7.2))
\begin{equation}\label{quartic}
	\sim \Lambda^4\int\dd[4]{x}\sqrt{g}
\end{equation}
and
\begin{equation}\label{quadratic}
	\,\,\,\,\,\,\,\sim \Lambda^2\int\dd[4]{x}\sqrt{g}\, R\,.
\end{equation}
Moreover, they observe that these terms, that depend quartically and quadratically on the running scale $\Lambda$, are at the origin of the UV-attractive fixed point of the asymptotic safety scenario. We will comment on the existence/absence of this fixed point at the end of the present section.

In \cite{Bonanno:2025xdg}, it is also claimed that the use of the FV measure in the path integral amounts to make in\,\eqref{quartic} and\,\eqref{quadratic} (that are obtained using the Fujikawa measure) the replacement $\Lambda^2\to N^2 g^{00}$, where $N$ is a dimensionless (running) cutoff. They then get
\begin{equation}\label{schif1}
	\,\,\,\,\Lambda^4\int\dd[4]{x}\sqrt{g}\quad\to\quad N^4\int\dd[4]{x}(g^{00})^2\sqrt{g}
\end{equation}
and
\begin{equation}\label{schif2}
	\Lambda^2\int\dd[4]{x}\sqrt{g}\, R\quad\to\quad N^2\int\dd[4]{x}\sqrt{g}\,g^{00}\, R\,.
\end{equation}
Accordingly, they claim that the FV measure leads to the presence of non-invariant operators in the running action $S_{\Lambda}[g]$. These claims are incorrect. In fact, from the thorough discussion above, it is clear that these claims are due to the loss of important terms in their calculation of $S_{\Lambda}[g]$.
We have shown that when the FV measure is used, not only non-invariant terms of the kind\,\eqref{schif1} and\,\eqref{schif2} do not appear, but also that the presence of $g^{00}$ factors in this measure is necessary to guarantee the diffeomorphism invariance of $S_{\Lambda}[g]$ (see\,\eqref{scalarres} and comments therein).
More in detail, we have seen that the $g^{00}$ factors in the FV measure turn out to cancel terms that arise when the different $\log(-\square^{(s)})$ in $\delta S_{\rm grav}$ are correctly calculated (here $-\square^{(s)}$ are the Laplace-Beltrami operators for the different spins, $s=0,1,2$), and only diffeomorphism invariant terms are left in the final result. Among them are the terms\footnote{In \cite{Branchina:2024lai}, the calculations are performed with a spherical background. Moreover, the dimensionless (running) cutoff $N$ is indicated with $L$. For the sake of clarity, here we uniform our notation to \cite{Bonanno:2025xdg} and use $N$.} $N^4/3$ and $-34N^2/3$ in our Eq.\,(30) of \cite{Branchina:2024lai}. Contrary to what is incorrectly claimed in \cite{Bonanno:2025xdg}, they are diffeomorphism invariant and
do not come from\,\eqref{schif1} and\,\eqref{schif2} evaluated in a coordinate system where $g^{00}=R$.

Let us finally go back to the question of the possible existence of the UV-attractive fixed point of the AS scenario. It is true that this fixed point would exist if terms of the kind\,\eqref{quartic} and\,\eqref{quadratic} appeared in the RG equation for the running action. However, thanks to a careful treatment of the path integral measure and a proper introduction of the physical running scale, in \cite{Branchina:2024lai} we show that a term like\,\eqref{quartic} is absent, and that (more generally) the beta functions for the Newton and the cosmological constant are significantly different from those of the AS literature \cite{Bonanno:2004sy,Reuter:2001ag}. They do not possess the non-trivial UV-attractive fixed point of the AS scenario. 

\section{Conclusions}
\label{Conclusions}

In the present paper we have thoroughly considered the controversial issue concerning the diffeomorphism invariance of the path integral measure in quantum gravity. It is usually thought that, due to the presence of non-covariant $g^{00}$ factors, the FV measure cannot be diffeomorphism invariant. 

With the help of the detailed analysis developed in sections \ref{diffinvmeas}, \ref{transfcoeff} and \ref{FVandFuji}, we have shown that, contrary to this na\"ive expectation, the $g^{00}$ factors that appear in the FV measure are necessary to ensure the diffeomorphism invariance of the effective action $\Gamma[g]$. Actually, we have seen that the invariance emerges from a delicate balance between {\it all} the elements involved in the definition of the path integral. In particular, we have shown that a crucial point is related to the necessity of introducing a time ordering parameter and a discretization (lattice) of spacetime. Differently from other gauge theories, where the gauge transformation does not affect the spacetime coordinates, when considering a general coordinate transformation $x\to\hat x$, two different lattices and time ordering parameters are involved in the two coordinate systems. As a consequence, non-trivial factors appear when the path integral undergoes a general coordinate transformation. The $g^{00}$ factors of the FV measure are necessary to {\it exactly} compensate for these non-trivial terms, and this is what ensures the diffeomorphism invariance of the effective action.  

Another important point concerns the derivation of the configuration space measure from the phase space one. The FV measure {\small $\prod_{ x}\big[(g^{00}(x))^{1/2}(g(x))^{1/4}\dd{\phi( x)}\big]$} is obtained after integration over the conjugate momenta $\pi$ if the phase space measure is the Liouville one, namely {\small $\prod_{ x}\big[\dd{\pi(x)}\dd{\phi(x)}\big]$}. The latter is what one would naturally expect from considerations based on the path integral construction of the basic transition amplitude $\braket{\phi',t'}{\phi'',t''}$. On the contrary, the Fujikawa measure is obtained in configuration space if {\small $\prod_x \big[(g^{00}(x))^{-1/2}\,\dd{\pi(x)}\dd{\phi(x)}\big]$} is assumed to be the phase space measure. This rather bizarre form of the phase space measure was first proposed in \cite{Toms:1986sh}, though the arguments presented in that paper are far from being well-grounded. This proposal was recently taken up by the authors of \cite{Bonanno:2025xdg}. Based on their idea that the diffeomorphism invariant measure in configuration space is the Fujikawa one, they derive this bizarre phase space measure from the request that, after integration over the conjugate fields $\pi$, the resulting configuration space measure is the Fujikawa one. The detailed analysis presented in this work, where we have shown that it is rather the FV measure to be invariant, indicates that such an artificial distortion of the natural Liouville measure in phase space has to be rejected (as one also expects on the basis of simpler arguments).

In conclusion, we have shown that the question of the measure to be used in the calculation of the effective gravitational action $\Gamma[g]$ presents delicate aspects that can be missed if one resorts to {\it formal} calculations. Paying attention to the construction of the path integral involved in the very definition of $\Gamma[g]$, it turns out that the measure proposed by Fradkin and Vilkovisky ensures the diffeomorphism invariance of $\Gamma[g]$, while the measure proposed by Fujikawa does not.

\section{Acknowledgements}
The work of CB has been supported
by the European Union – Next Generation EU through the research grant number P2022Z4P4B
“SOPHYA - Sustainable Optimised PHYsics Algorithms: fundamental physics to build an advanced
society” under the program PRIN 2022 PNRR of the Italian Ministero dell’Università e Ricerca
(MUR). The work of VB, FC, RG and AP is carried out within the INFN project QGSKY.

\end{document}